\documentclass[pra,showpacs]{revtex4}
\usepackage{amssymb}
\usepackage[dvips]{graphicx}
\usepackage{amsmath}
\usepackage{graphicx}
\usepackage{makeidx}
\usepackage{subfigure}

\begin{document}
\title{Kinetic Friction  by a Small Number of Intervening Inelastic Particles between Rough Surfaces} 
\author{Hidetsugu Sakaguchi}
\affiliation{Department of Applied Science for Electronics and Materials,
Interdisciplinary Graduate School of Engineering Sciences, 
Kyushu University, Kasuga, Fukuoka, Japan 816-8580}
\begin{abstract}
We investigate a mechanism of the appearance of kinetic friction in granular materials. We consider a small number of intervening inelastic particles between two rough surfaces as one of the simplest dynamical models to study granular friction.  The resistance force applied to the upper surface is numerically calculated. We find that the resistance force $F(t)$ is scaled as $F^{\prime}(vt)$ for a small pulling velocity $v$.  The time average $F_0=\langle F(t)\rangle$ in the limit $v\rightarrow 0$ is not zero owing to the mutual collisions between the intervening particles. The nonzero $F_0$ implies the appearance of kinetic friction in this simple dynamical system. 
\end{abstract}
\maketitle
\section{Introduction}
 Friction forces have been long studied since the formulation of Coulomb-Amonton's laws. The laws of friction  have been widely studied from the macroscopic scale to the nano scale~\cite{rf:1,rf:2,rf:3,rf:4}.  The establishment of frictional contact between two solids by a normal force and its disruption by a tangential force are important processes in friction, and the frictional contact has been directly observed in experiments~\cite{rf:5}.
Recently, M\"user and coworkers and Daly et al. have proposed another mechanism of friction, that is, the origin of Coulomb-Amonton's laws is the presence of intervening dirt molecules or atoms  between two solid surfaces~\cite{rf:6,rf:7,rf:8}.  In a model of  kinetic friction~\cite{rf:7}, M\"user assumed a potential between the intervening particle and the top (or bottom) surface such as $V_{t,b}=V^0_{t,b}\cos(x/b_{t,b})+V^1_{t,b}\cos(2x/b_{t,b})+\cdots$, where $2\pi b_{t,b}$ are the periods of the top and bottom wall, respectively.  A certain viscous force acting on the lubricant atom is further assumed for energy dissipation. A smooth motion is observed for the intervening particle and no kinetic friction appears, if the interaction force between the particle and the surface is a simple sinusoidal one, i.e., $V^n_{t,b}=0$ for $n\ge 1$; however, kinetic friction appears as a result of a stick-slip motion, if the interaction forces include a certain  higher-harmonic component, e.g., $V^0_{t,b}>0,V^1_{t,b}>0$, and $V^n_{t,b}=0$ for $n\ge 2$.  

Recently, the friction and rheological properties of granular materials have been intensively studied.  Granular materials exhibit various states such as a gaslike state, a liquidlike state and a solidlike state. They are numerically studied by molecular dynamics simulations including many inelastic particles. One of the interesting phenomena is the clustering in the gas phase owing to inelastic collisions~\cite{rf:9}.  
Stick-slip motions caused by the frictional vibration  have been studied numerically and experimentally~\cite{rf:10,rf:11,rf:12}. They are related to 
the transition  between the liquidlike state and the solidlike state~\cite{rf:13,rf:14}. The transition from the liquidlike state to the solidlike state when the density is increased is called jamming transition~\cite{rf:15}. Several scaling relations were found for the static properties such as the pressure and the average coordination number near the jamming transition~\cite{rf:16}. Scaling laws for rheological properties of sheared granular materials were also studied by numerical simulations~\cite{rf:17,rf:18}.  Below the critical density, the shear stress $S$ is proportional to the square of shear velocity. This is called the Bagnold scaling~\cite{rf:19}. 
Above the critical density, $S$ is not zero even for an infinitesimal shear rate. The nonzero shear stress indicates the rigidity of the solid phase. A nonzero shear stress can be interpreted as the appearance of a kinetic friction in the granular materials. The nonzero shear stress appears even in an assembly of frictionless inelastic particles.  That is, even if there is no friction among elemental particles, macroscopic friction appears.  A force-chain network appears above the critical density in granular materials. It has been suggested that the anisotropy of the force-chain network is the origin of kinetic friction~\cite{rf:20}. However, the origin of kinetic friction is not well understood, even in an assembly of such simple frictionless inelastic particles. A granular material begins to flow when the shear stress is beyond its critical value, called the yield stress in experiments, under a constant shear stress. The existence of yield stress is related to the appearance of the maximum static friction. 

In this study, we will mainly analyze  kinetic friction by numerical simulation of a small number of inelastic intervening particles, expecting that some aspects of the friction may be clarified, if the degree of freedom of motion is reduced. Firstly, we neglect tangential forces such as the Mindlin force and rotational motion to simplify the system as much as possible.  A small number of frictionless inelastic particles are intervening between the upper and lower walls. 
Our model is related to the M\"user model in that the intervening particles play  an important role, but the model equation is that conventionally used in the numerical simulation of granular materials. The upper and lower walls are rough surfaces constructed from an array of semicircles. Inelastic collisions occur among the intervening particles and between the intervening particles and the rough surfaces. We numerically calculate the resistance force exerted on the upper wall by the intervening particles, and discuss the origin of kinetic friction. In \S 2, we will discuss kinetic friction, static friction, and stick-slip motion in a single-particle system.  In \S 3, we will study a two-particle system, and point out the importance of the mutual collisions between intervening particles for the appearance of kinetic friction. In actual experiments, the rotational degree of motion plays an important role~\cite{rf:12,rf:21}. In \S 4, we incorporate rotational motion by adding some tangential forces, and discuss several effects of rotational motion.  In this study, no explicit form of friction is assumed between elemental particles.

\section{Resistance Force by a Single Intervening Particle}   
We first consider a single-particle system confined between two walls. 
The particle is assumed to be a circle of radius $r=r_1$. The upper and lower walls are constructed from periodic arrays of semicircles of radius $r=r_2=0.5\ge r_1$ as shown in Fig.~1(a). The $y$-coordinates of the centers of the semicircles in the lower and upper walls are respectively fixed to be 0 and $h>0$.
We assume that the particle and semi-circles interact with a linear elastic force and a viscous force, only if the distance $|{\bf x}_j-{\bf x}_i|$ between the particle $i$ and the semicircle $j$ is smaller than the sum of the radii of the two particles.   Then, the intervening particle obeys the equation of motion;
\begin{equation}
m\frac{d^2{\bf x}_i}{dt^2}=\sum_{j}\{K(r_j+r_i-|{\bf x}_j-{\bf x}_i|)+\eta ({\bf v}_j-{\bf v}_i)\cdot {\bf n}_{ij}\}{\bf n}_{ij}\delta_{ij},
\end{equation}
where $i=1$ indicates the particle intervening between the two walls, $m$ is the mass of the particle, $\delta_{ij}=1$ if $|{\bf x}_j-{\bf x}_i|\le r_j+r_i$, $\delta_{ij}=0$ if $|{\bf x}_j-{\bf x}_i|>r_j+r_i$, ${\bf n}_{ij}=({\bf x}_j-{\bf x}_i)/|{\bf x}_j-{\bf x}_i|$, and ${\bf v}_j=d{\bf x}_j/dt$. 
A summation is taken for all the semicircles in contact with the particle.
The reaction force exerted on the semicircles in the upper and lower walls by the particle is calculated to study friction force. The parameter values are $r_1=0.425$, $m=4r_1^2$,$K=0.42\times 10^4$, and $\eta=83.9$. A short-range viscous force effectively plays a role of an inelastic collision. Periodic boundary conditions are assumed at $x=1$ and $x=2$, that is, the position of the particle is reset to $x=1$ when $x$ reaches the right boundary $x=2$. Tangential forces can be assumed between two particles as elemental forces, which is more realistic for granular particles. However, we neglect such forces in this section for simplicity. 

Firstly, we performed a type of numerical simulation in which the upper wall is pulled at a constant velocity $v$ in the $x$-direction, and the lower wall is fixed. The resistance force $F$ exerted on the upper wall by the intervening particle was numerically evaluated. Figure 1(b) shows the time evolution of $F$ at$v=0.02$ for $h=0.56$. The resistance force $F(t)$ changes periodically with time.  
Figure 1(c) shows the time average $\langle F\rangle$ of the resistance force $F(t)$ as a function of $v$ for $h=1.68,1.64$, and $1.6$. The average resistance force is roughly proportional to $v$ for any $h<h_c\sim 1.703$. 
The intervening particle is trapped owing to the inelastic collisions by the lower wall, and the average velocity becomes 0 for $h<h_c$.  This is partially related to the assumption that the rotational degree of freedom is neglected. 
The unidirectional motion induced by tangential forces is discussed later in \S4. At the critical height, a semicircle in the upper wall is located just above the intervening particle, and the particle is located just between the two semicircles in the lower wall, as shown in Fig.~1(a). 
The centers of the two semicircles of radius $r_2$ are denoted as A and C, and the center of the particle of radius $r_1$ is denoted as B;  D is the contact point of the two semicircles in the lower wall in Fig.~1(a). 
The critical height can be evaluated as $h_c=r_1+r_2+\sqrt{(r_1+r_1)^2-r_2^2}\sim 1.703$, because AB$=r_1+r_2$, BD$=\sqrt{(r_1+r_1)^2-r_2^2}$, and $h_c=$AD. 

We have also numerically calculated the normal force $N$ applied to the upper wall by the intervening particle. The time average of the normal force $N$ increases as $h$ decreases from $h_c$.  Figure 1(d) shows $\mu=\langle F\rangle/\langle N\rangle$ for $h=1.68,1.64$, and $1.6$. The coefficients $\mu$ overlap rather well in a range of  small $v$ values for three different values of $h$, and $\mu$ approaches zero as $v\rightarrow 0$. Figure 1(e) shows the proportional constant $\nu=\langle F\rangle /v$ for a small $v$  as a function of $h$. The proportional constant $\nu$ increases as $h$ decreases. 
\begin{figure}[tbp]
\includegraphics[width=12.cm]{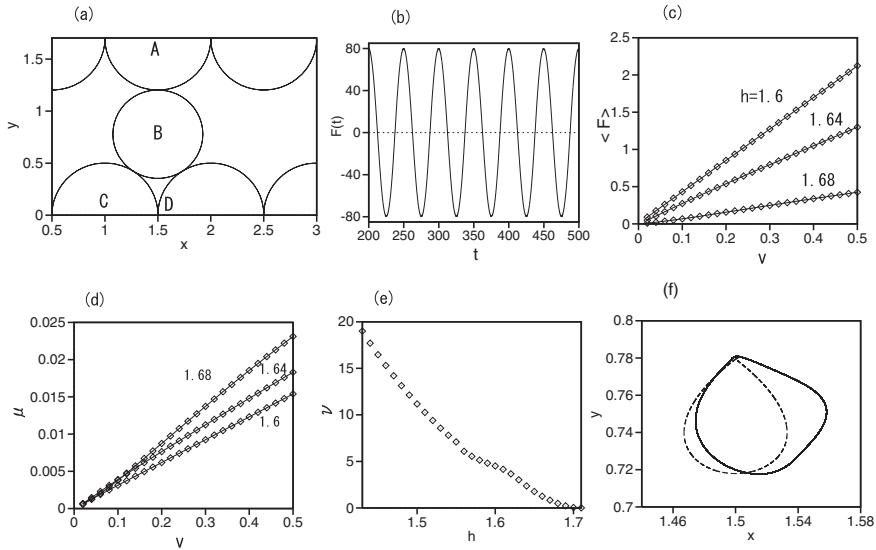}
\caption{(a) A particle is intervening between the upper and lower walls. The height $h$ between the two walls is 1.703. (b) Time evolution of $F(t)$ at $v=0.02$ for $h=1.56$. (c)  Time average $\langle F\rangle$ of the resistance force as a function of $v$ for $h=1.69,1.64$, and $1.59$. 
(d) Ratio $\mu=\langle F\rangle/\langle N\rangle$ as a function of $v$ for $h=1.69,1.64$, and $1.59$. (e) Proportional constant $\nu=\langle F\rangle/v$ for a small $v$ as a function of $h$. (f) Trajectories of the particle during the slip process at $v=0.01$ (dashed curve) and $4$ (solid curve) for $h=1.56$.}
\label{fig1}
\end{figure}

\begin{figure}[tbp]
\includegraphics[width=12.cm]{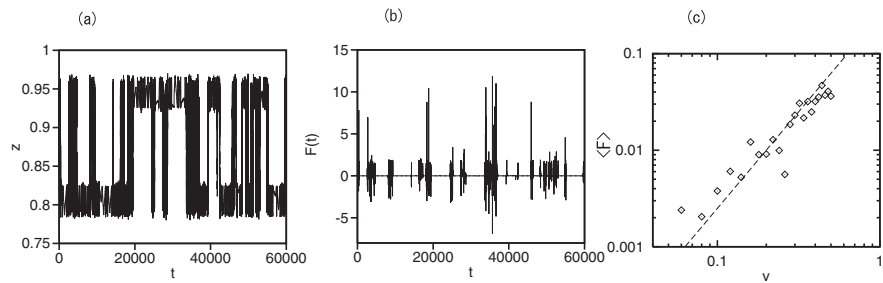}
\caption{(a) Time evolution of  $z$ of the intervening particle at $h=1.75$. (b)  Time evolution of $F(t)$ at $h=1.75$. (c) Double-logarithmic plot of time average $\langle F\rangle$ as a function of $v$. The dashed line denotes $\langle F\rangle \propto v^2$.}
\label{fig2}
\end{figure}

\begin{figure}[tbp]
\includegraphics[width=9.cm]{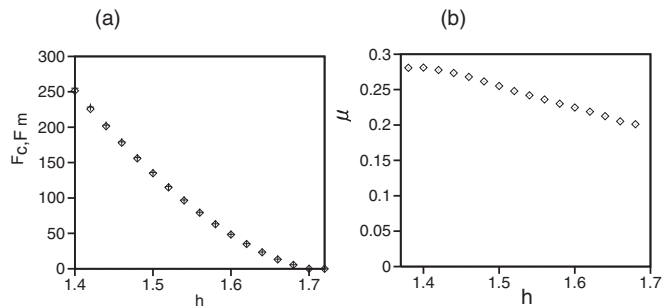}
\caption{(a) Maximum static friction $F_c$ denoted by $\diamond$ in constant-force simulations, and the maximum value $F_m$ denoted by $+$ of the time-periodic resistance force in constant-velocity simulations.  
(b) Coefficient of the maximum static friction as a function of $h$.}
\label{fig3}
\end{figure}
\begin{figure}[tbp]
\includegraphics[width=9.cm]{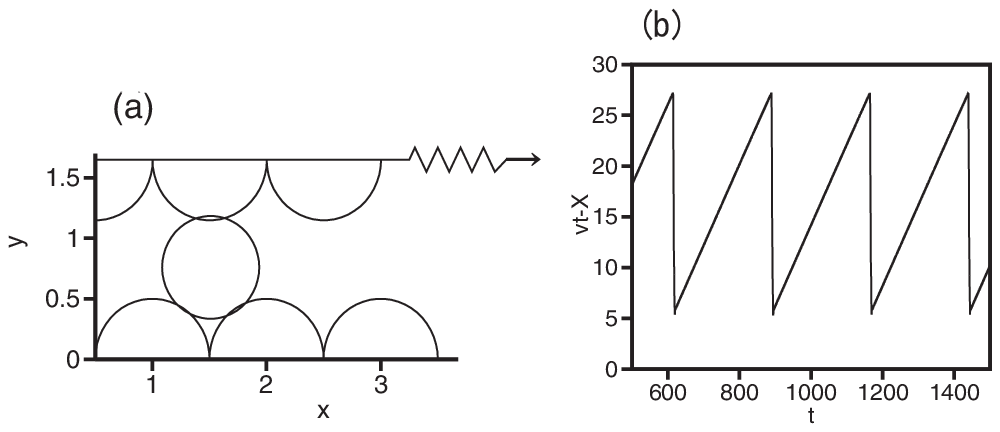}
\caption{
(a)  One-particle system with the upper wall coupled to a spring. The end point of the spring is pulled at a constant velocity $v$.
(b) Time evolution of $vt-x$ of the upper wall. A stick-slip motion appears.}
\label{fig4}
\end{figure}

The intervening particle is trapped between the two semicircles on the lower wall for $h<h_c$, and the upper wall slips on the particle. The average velocity of the intervening particle is therefore zero. 
From the Galilean symmetry, it is possible that the average velocity of the intervening particle is $v$ and the lower wall slips below the moving particle. These two slip states are bistable for $h<h_c$. Figure 1(f) shows two trajectories of the particle during the slip process at $v=0.005$ and $3.5$ for $h=1.56$. At $v\sim 0$, the trajectory is almost symmetric with respect to $x=1.5$, and the average value of the resistance force $F$ is almost zero from the symmetry. As $v$ is increased, the $x$-coordinate of the trajectory shifts toward the right,  then the average $\langle F\rangle$ becomes nonzero as a result of the asymmetric trajectory of the particle. Generally, the kinetic friction on a solid surface takes a nonzero value, even if $v$ is sufficiently small. Thus, we conclude that no kinetic friction appears in this single-particle system.

This single-particle system is a very simple dynamical system, but its dynamics is rather complex. If the interval $h$ between the two walls is larger than the critical value $h_c$, it is possible that the intervening particle is trapped in the lower wall (or the upper wall) and that the particle is never in contact with the upper wall (or the lower wall).  Then, the resistance force is completely zero.  For general initial conditions, the intervening particle collides with the two walls in a chaotic manner. This might correspond to the phase of granular gas or granular liquid.  Figure 2(a) shows the time evolution of the vertical coordinate $z(t)$ of the intervening particle at $h=1.75$.  The intervening particle is located near $z\sim 0.8$ and $z\sim 0.95$ for most of the time. This corresponds to the two trapping states in the lower wall ($z\sim 0.8$) and upper wall ($z\sim 0.95$). The switching between the two trapping states occurs intermittently. Figure 2(b) shows the time evolution of the resistance force $F(t)$ at $h=1.75$.  The time evolution is also intermittent. 
The resistance force $|F(t)|$ takes a large value during the switching periods. Figure 2(c) shows a logarithmic plot of the long time average of $F(t)$ as a function of $v$ for $h=1.75$. The solid line denotes $\langle F\rangle\propto v^2$, which might correspond to the Bagnold scaling found in a numerical simulation of granular materials~\cite{rf:18}.  

In most experiments on friction, a constant external force is applied to a solid body, and in some experiments, a solid body is pulled at a constant velocity via a spring.  We can perform  corresponding  numerical simulations in our single-particle system. That is, the upper wall is pulled in the $x$-direction by a constant external force $F_{ex}$. The upper wall moves as a rigid body. We have calculated the resistance force $F(t)$ exerted on the upper wall by the intervening particle.  If the external force is small, a stationary state is finally obtained. 
That is, the resistance force $F$ is equal to $F_{ex}$ and the upper wall and the intervening particle do not move in the final stationary state.  
If $F_{ex}$ is larger than its critical value $F_c$, the upper wall begins to move toward the right. The critical force $F_c$ is interpreted as the maximum static friction. The rhombi in Fig.~3(a) show the maximum static friction as a function of $h$.  The maximum static friction decreases monotonically with $h$. 
In numerical simulations of constant velocity, the resistance force $F(t)$ changes periodically with time at $h<h_c$ as shown in Fig.~1(b).
We calculated the maximum value $F_m$ of the time-periodic resistance force $F(t)$ as a function of $h$.  The maximum value $F_m$ for $v=0.02$ is shown by $+$ in Fig.~3(a). It is natural that the maximum static friction $F_c$ is equal to $F_m$.  Figure 3(b) shows the coefficient $\mu=F_c/N$ of the maximum static friction as a function of $h$. The coefficient $\mu$ is a slightly decreasing function of $h$.  The coefficient of the maximum static friction is not zero even in this single-particle system.  The maximum static friction corresponds to the existence of yield stress in granular materials. 
 
We also performed another type of numerical simulation. The upper wall is forced via a spring whose right end is pulled at a constant velocity $v$, as shown in Fig.~4(a). The center of mass of the upper wall obeys 
\begin{equation}
M\frac{d^2X}{dt^2}=\sum_{i,j}K(r_j+r_i-|{\bf x}_j-{\bf x}_i|)+\eta ({\bf v}_j-{\bf v}_i)\cdot {\bf n}_{ij}\}({-\bf n}_{ij}\cdot {\bf n}_x)\delta_{ij}+k(vt-X),
\end{equation}
where $X$ is the $x$-coordinate of the center of mass of the upper wall, ${\bf n}_x$ is a unit vector in the $x$-direction, $M$ is the mass of the upper wall, and $k$ is the spring constant.  Figure 4(b) shows the time evolution of $vt-X$ for $M=5,k=5$, and $v=0.08$ at $h=1.5$. A stick-slip motion is clearly seen.  The upper wall almost sticks to the intervening particle and the lower wall for most of the time owing to the friction. The upper wall slips forward abruptly by $\Delta X \sim 21$ in each period, which is much larger than the diameter  of semicircles. The oscillation is not due to the periodic force given by the array of the semicircles, but is a friction-induced vibration observed in many frictional systems. Such frictional vibration occurs because the maximum static friction is larger than the kinetic friction, or the frictional force becomes weaker when the solid body begins to move. In our system, the kinetic friction is effectively zero; therefore, it is natural to observe the frictional vibration. This type of stick-slip motion is  observed experimentally in granular systems.~\cite{rf:10,rf:11} 
\section{Kinetic Friction by Intervening Particles}
As shown in the previous section, no kinetic friction appears in a single-particle system. We consider a two-particle system in this section. To investigate  kinetic friction, the upper wall is pulled at a constant velocity $v$.   Two particles with the same radius $r=r_1=0.425$ are intervening between the upper and lower walls. 
The periodic boundary conditions are imposed at $x=1$ and $x=3$ for this two-particle system. In the two-particle system, the intervening particles interact with each 
other with the same types of elastic and viscous forces, only when the distance between the two particles is smaller than $2r_1$. The two intervening particles also interact with the upper and lower walls. 
The initial conditions are $x_1=1.05,y_1=h/2,v_{x1}=v_{y1}=-0.05,x_2=2.102,y_2=h/2$, and $v_{x2}=v_{y2}=-0.05$ for the two intervening particles. The initial conditions for the semicircles in the lower wall are  $x_j=j,y_j=0$, and $v_{xj}=0$.  For the semicircles in the upper wall, $x_j=j+0.2505,y_j=h$, and $v_{xj}=v=0.01$.  The velocity $v$ of the upper wall is increased stepwise to $0.01\times (n+1)$ at $t=2500\times n$ ($n=1,2,\cdots$).  There is another critical width $h_{c2}\sim 1.55$ in this two-particle system. For $h_{c2}<h<h_c$, the average resistance force $\langle F\rangle$ per particle is roughly proportional to $v$, which is similar to the behavior of the single-particle system. This is because the two particles are almost trapped between different neighboring semicircles in the lower wall, and there is no interaction between the two intervening particles. However, the behavior of the resistance force changes when $h$ is smaller than $h_{c2}$. Figure 5(a) shows the average resistance force $\langle F\rangle$ as a function of $v$ for $h=1.56,1.51$, and $1.46$ for the two-particle system. At $h=1.56>h_{c2}$, the average resistance force is roughly proportional to the velocity $v$. At $h=1.51$ and $1.46$, the average resistance force takes a nonzero value, $F_0$, even for a sufficiently small $v$. The nonzero value $F_0$ in the limit  $v\rightarrow 0$ appears when $h$ is smaller than $h_{c2}\sim 1.55$. We can interpret this resistance force as a kind of kinetic friction. Figure 5(b) shows $F_0$ as a function of $h$. 
We have calculated the normal force acting on the upper wall and evaluated the kinetic friction coefficient $\mu_s$ for a sufficiently small $v$. Figure 5(c) shows the coefficient $\mu_s$ as a function of $h$. On the other hand, the coefficient of the maximum static friction for the two-particle system was the same as that in Fig.~3(b). The kinetic friction coefficient is definitely smaller than the coefficient of the maximum static friction in our system.

\begin{figure}[tbp]
\begin{center}
\includegraphics[width=12.cm]{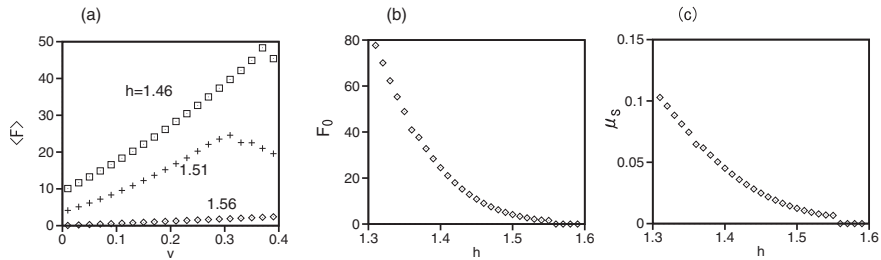}
\caption{(a) Resistance force $\langle F\rangle$ as a function of $v$ for $h=1.56,1.51$, and $1.46$. 
(b) Kinetic friction force $F_0$ for infinitesimally small $v$ as a function of $h$.
(c) The coefficient of the kinetic friction as a function of $h$}
\end{center}
\label{fig5}
\end{figure}
\begin{figure}[tbp]
\begin{center}
\includegraphics[width=12.cm]{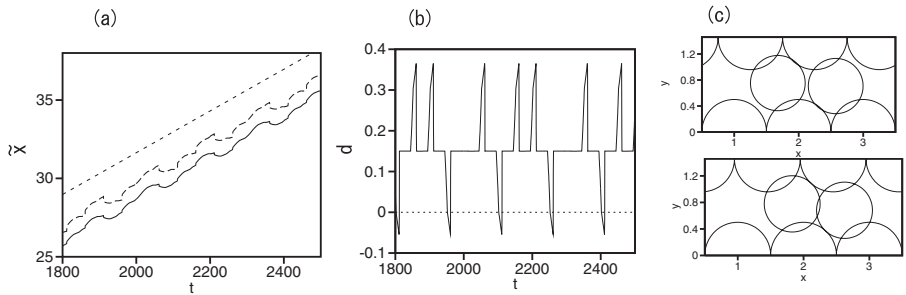}
\caption{(a) Time evolutions of $\tilde{x}_1(t)$ and $\tilde{x}_2(t)$ for $h=1.46$ and $v=0.02$. (b) Time evolution of $d=\sqrt{(x_1(t)-x_2(t))^2+(y_1(t)-y_2(t))^2}-2r_1$ for $h=1.46$ and $v=0.02$. (c) Snapshot patterns for $h=1.46$ and $v=0.02$ at $t=2400$ (top) and $t=2410$ (bottom). }
\end{center}
\label{fig6}
\end{figure}
\begin{figure}[tbp]
\begin{center}
\includegraphics[width=8.cm]{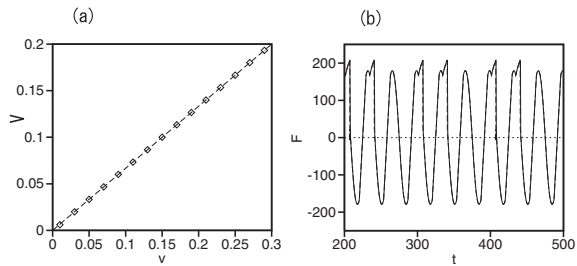}
\caption{(a) Average velocity $V$ of $\tilde{x}_1$ as a function of $v$. (b) Time evolution of the resistance force $F$ at $v=0.03$ (solid curve) and $F^{\prime}=F(t/3+200/3)$ at $v=0.01$ (dashed curve). The difference is hardly observed.}
\end{center}
\label{fig7}
\end{figure}
\begin{figure}[tbp]
\begin{center}
\includegraphics[width=11.cm]{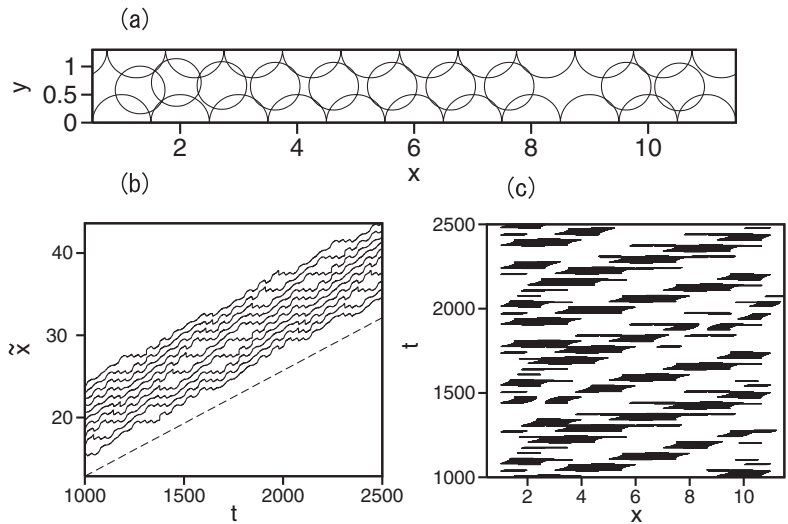}
\caption{(a) Snapshot pattern of a ten-particle system at $t=2500$ for $v=0.03$ and $h=1.3$. (b) Time evolution of $\tilde{x}_i$ for $i=1,2,\cdots, 10$ at $v=0.03$ and $h=1.3$. The dashed line denotes $\tilde{x}\sim (3/7)vt$. (c) Spatio-temporal plot of mutual collisions. Neighboring particles satisfying $d=\sqrt{(x_1(t)-x_2(t))^2+(y_1(t)-y_2(t)^2}-2r_1<0$ at time $t$ are linked with a line.}
\end{center}
\label{fig8}
\end{figure}
The appearance of a nonzero value $F_0$ is closely related to the mutual collisions of the two intervening particles.  
Figure 6(a) shows the time evolutions of $\tilde{x}_1(t)$ and $\tilde{x}_2(t)$ for $h=1.46$ and $v=0.02$, where $\tilde{x}_{1}$ (solid curve) and $\tilde{x}_2$  (dashed curve) denote the $x$-coordinates for the first and second particles correspondingly; however, no resetting process $x\rightarrow x-2$ under the periodic boundary conditions is taken into account in the notation $\tilde{x}$. The dotted line represents $\tilde{x}\sim 0.01333t$. The intervening particles are moving at an average velocity $0.0133$, which is different from the velocities of 0.02 and 0 of the upper and lower walls.  Figure 6(b) shows the time evolution of $d=\sqrt{(x_1(t)-x_2(t))^2+(y_1(t)-y_2(t))^2}-2r_1$. Mutual collision occurs when $d$ is negative.   Figure 6(c) shows two snapshots of the two particles at $t=2400$ and $2410$. The mutual collision is observed in the snapshot at $t=2410$. 

Figure 7(a) shows the average velocity $V$ of $\tilde{x}_1$ as a function of $v$. The average velocity satisfies $V=(2/3)v$.  The ratio of the velocity of the intervening particles to that of the upper wall is a simple fraction, which implies a type of locking phenomenon. The locking phenomenon appears because of the spatially periodic structure of the upper and lower walls. 
These locking behaviors depend on the initial conditions. 
That is, there are multistable states in this system. 
The trapping states with $V=0$ or $V=v$ are also stable states even for $h<h_{c2}$. We have confirmed the existence of other locking states with $V=(1/2)v$ and $V=(5/6)v$ for the same $h$. 

To understand the appearance of the nonzero value $F_0$ for a sufficiently small $v$, we have investigated the time evolution of the resistance force $F(t)$ applied to the upper wall in more detail. The solid curve in Fig.~7(b) shows the time evolution of $F(t)$ at $v=0.03$. The resistance force changes periodically with time, and takes positive and negative values. The behavior is similar to that in the case of the single-particle system. However, at the moment of mutual collisions, impulsive forces are exerted on the lower and upper walls by the intervening particles. In the time evolution of $F(t)$ in Fig.~7(b), the impulsive forces  appear as the second peak in the doublet peaks in the range of $F>0$. As a result of the additional impulses, the time average of $F(t)$ takes a positive value $\langle F\rangle=11.7$ at $v=0.03$.  
The dashed curve in Fig.~7(b)  shows the time evolution of the resistance force at $v=0.01$. However, the time scale is reduced by one-third. That is, $F^{\prime}=F(t/3+200/3)$ is shown in this figure. The solid and dashed curves are well overlapped. This implies that the time evolution is scaled as $F(t)=F^{\prime}(vt)$ for a sufficiently small $v$, and therefore, the time average  $\langle F\rangle$ of $F$ is nonzero even if $v$ is sufficiently small. 
That is, a certain number of impulsive forces are applied to the upper wall, while the upper wall shifts by one period in the $x$-direction. For a sufficiently small $v$, the process occurs slowly, but the total force applied to the upper wall per spatial period does not depend on the velocity $v$.    
This is the origin of the appearance of a nonzero $F_0$ in our two-particle system.       
 
Furthermore, we have studied kinetic friction in some many-particle systems. Figure 8 shows some numerical results for a ten-particle system at $h=1.3$.
Figure 8(a) displays a snapshot pattern of the ten particles at $t=2500$ for $v=0.03$. Three intervening particles located near $x\sim 2$ are in contact with each other in the snapshot. 
Figure 8(b) shows the time evolution of $\tilde{x}_i$ for $i=1,2,\cdots, 10$. All intervening particles move at the same average velocity $V\sim (3/7)v$. 
This is another example of the locking phenomenon.  
Figure 8(c) shows a spatio-temporal plot of mutual collisions. That is, two particles satisfying $d=\sqrt{(x_1(t)-x_2(t))^2+(y_1(t)-y_2(t)^2}-2r_1<0$ are linked by a line at time $t$. Clusters of two, three, or four intervening particles are created and annihilated with time by mutual collisions. The mutual collisions occur sequentially, and they propagate in the $x$-direction.  The nonzero resistance force $F_0$ is also observed in this system. The impulsive forces induced by the mutual collisions between intervening particles might be the origin of kinetic friction. 

\section{Kinetic Friction by Intervening Particles with Rotational Degrees of Freedom}
In the previous sections, we have neglected tangential forces. If the tangential forces are neglected, no torque is applied to each particle and then the rotational degree of freedom can be neglected.  In this section, we incorporate a tangential elastic force and a tangential viscous force.  The equation of motion of the intervening particle is written as 
\begin{eqnarray}
m\frac{d^2{\bf x}_i}{dt^2}&=&\sum_{j}[\{K(r_j+r_i-|{\bf x}_j-{\bf x}_i|)+\eta ({\bf v}_j-{\bf v}_i)\cdot {\bf n}_{ij}\}{\bf n}_{ij}\nonumber\\&&+\{K_t\Delta(({\bf r}_j-{\bf r}_i)\cdot {\bf t}_{ij}+r_i\theta_i+r_j\theta_j)+\eta_t (({\bf v}_j-{\bf v}_i)\cdot {\bf t}_{ij}+r_i\omega_i+r_j\omega_j)\}{\bf t}_{ij}]\delta_{ij},\nonumber\\
\end{eqnarray}
where  $K_t$ is the elastic constant for the tangential displacement, $\theta_i$ is the rotation angle of the $i$th particle, $\Delta(({\bf r}_j-{\bf r}_i)\cdot {\bf t}_{ij}+r_i\theta_i+r_j\theta_j)$ denotes the tangential displacement from the instance of contact between two particles, $\omega_i=d\theta_i/dt$, $\delta_{ij}=1$ if $|{\bf x}_j-{\bf x}_i|\le r_j+r_i$, $\delta_{ij}=0$ if $|{\bf x}_j-{\bf x}_i|>r_j+r_i$, and ${\bf t}_{ij}$ is a unit vector perpendicular to ${\bf n}_{i,j}$. 
The angle $\theta_i$ and $\omega_i$ for the semicircles in the upper and lower walls are set to be 0, because the lower wall is stationary and the upper wall is shifted at a constant velocity. If the rotational degree of freedom is taken into consideration,  the equation of rotation is written as
\begin{equation}
I\frac{d^2\theta_i}{dt^2}=\sum_{j}(-r_i)\times\{K_t\Delta(({\bf r}_j-{\bf r}_i)\cdot {\bf t}_{ij}+r_i\theta_i+r_j\theta_j)+\eta_t (({\bf v}_j-{\bf v}_i)\cdot {\bf t}_{ij}+r_i\omega_i+r_j\omega_j)\}\delta_{ij},
\end{equation}
where  $I$ is the inertial moment of the $i$th particle. The inertial moment is assumed to be $I=(1/2)mr_i^2$ in this simulation.  
\begin{figure}[tbp]
\begin{center}
\includegraphics[width=11.cm]{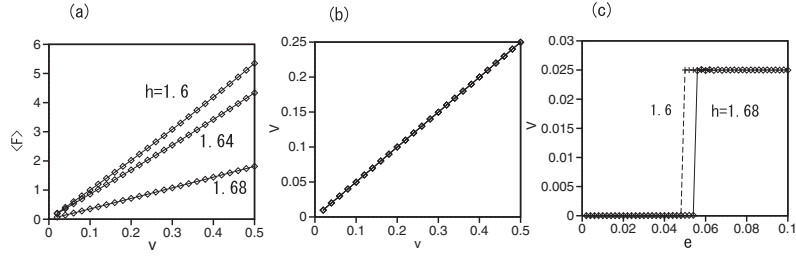}
\caption{(a) Average resistance force $\langle F\rangle$ for a single-particle system as a function of the velocity $v$ of the upper wall at $h=1.68,1.64$, and 1.6 for $\eta_t=0.1\eta$, and $K_t=0$. (b) Average velocity $V$ as a function of the velocity $v$ of the upper wall at $h=1.68,1.64$, and 1.6. (c) Average velocity $V$ of the intervening particle as a function of $e=\eta_t/\eta$ at $v=0.05$ for $h=1.68$ and 1.6. The parameter $e$ is increased stepwise by $\Delta e=0.0005$ from $e=0$.}
\end{center}
\label{fig9}
\end{figure}
Firstly, we neglect the tangential elastic force, that is, $K_t=0$.
Figure 9(a) shows the relation of the average resistance force $\langle F\rangle$ for a single-particle system and the velocity $v$ of the upper wall at $h=1.68,1.64$,  and 1.6 for $\eta_t=0.1\eta$. Figure 9(a) corresponds to Fig.~1(c) for $\eta_t=0$; however, the average resistance force $F$ is larger than that in the case of $\eta_t=0$, probably because of the additional viscous force. However, the resistance force $\langle F\rangle$ becomes zero when $v$ approaches 0. That is, no kinetic friction appears in the single-particle system, even if rotational motion is taken into consideration. Figure 9(b) shows the average velocity $V$ of the intervening particle as a function of the velocity $v$ of the upper wall. The average velocity $V$ satisfies $V=v/2$ for $\eta_t=0.1\eta$. For the  case of $\eta_t=0$, the velocity of the intervening particle is zero, that is, the intervening particle is trapped in the lower wall.  The tangential viscous force induces the flow of the intervening particle. Figure 9(c) shows the average velocity $V$ of the intervening particle as a function of $e=\eta_t/\eta$ at $v=0.05$ for $h=1.68$ and 1.6, when $e$ is increased stepwise by $\Delta e=0.0005$ from 0. A discontinuous transition from the trapped state to the flowing state is observed. The critical values $e_c$ are $0.056$ for $h=1.68$, $0.0535$ for $h=1.64$, and $0.05$ for $h=1.6$. When $e$ is decreased stepwise by $\Delta e=0.0005$ from 0.1, another discontinuous transition is observed.  The critical values $e_{c2}$ are $0.0555$ for $h=1.68$, $0.053$ for $h=1.64$, and $0.035$ for $h=1.6$. The transitions are almost discontinuous for $h=1.68$ and 1.64, but a hysteresis is observed for $h=1.6$. 
\begin{figure}[tbp]
\begin{center}
\includegraphics[width=9.cm]{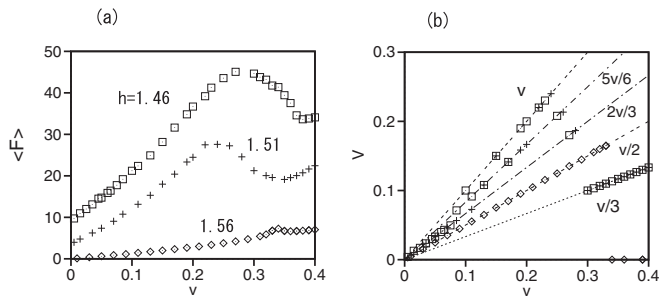}
\caption{(a) Average resistance force $\langle F\rangle$ for a two-particle system as a function of the velocity $v$ of the upper wall at $h=1.56,1.51$, and 1.46 for $\eta_t=0.1\eta$. (b) Average velocity $V$ of the first particle as a function of the velocity $v$ of the upper wall at $h=1.56,1.51$, and 1.46 for $\eta_t=0.1\eta$.}
\end{center}
\label{fig10}
\end{figure}
\begin{figure}[tbp]
\begin{center}
\includegraphics[width=9.cm]{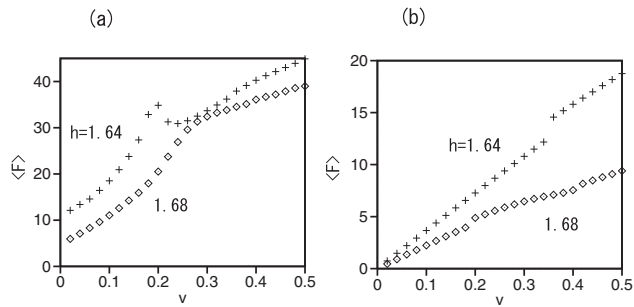}
\caption{(a) Average resistance force $\langle F\rangle$ for a single-particle system as a function of the velocity $v$ at $h=1.64$ and 1.68 for $K_t=0.01K$ and $\eta_t=\eta$. The rotational degree of freedom is not taken into consideration.
 (b) The same as (a), but the rotational degree of freedom is taken into consideration.}
\end{center}
\label{fig11}
\end{figure}
Figure 10 shows numerical results for a two-particle system at $\eta_t=0.1\eta$. Figure 10(a) shows the average resistance force $F$ as a function of $v$ for $h=1.46,1.51$, and 1.56. Figure 10(b) shows the average velocity $V$ of the first intervening particle as a function of $v$ for $h=1.46,1.51$, and 1.56. 
The average resistance force $\langle F\rangle$ becomes zero for $v\rightarrow 0$ at $h=1.56$. The average velocity is $V=v/2$ for $v<0.34$. At $h=1.56$, no mutual collisions between the intervening particles occur, and each particle behaves independently. However, a nonzero average resistance force appears for $h=1.46$ and 1.51. This implies the existence of the kinetic friction induced by mutual collisions between the intervening particles. The average velocity also satisfies $V=2v/3$ for a small $v$.  The relation $V=2v/3$ is not satisfied for a relatively large $v$ at $h=1.46$ and $h=1.51$, but other ratios such as $V/v=1,5/6$, and $1/3$ are observed. It might be an origin of the non-monotonic relation of $\langle F\rangle$ and $h$. 

Finally, we consider the effect of the tangential elastic force for a single-particle system.  First, the rotational degree of freedom is not taken into consideration in the numerical simulation, that is, $\theta_i$ is set to be zero at any $t$. Figure 11(a) shows the average resistance force $\langle F\rangle$ as a function of $v$ at $h=1.68$ and 1.64 for $K_t=0.01K$ and $\eta_t=\eta$.  The appearance of a nozero $F_0$ in the limit $v\rightarrow  0$  is seen in this figure. The nonzero $F_0$ increases with $K_t$. The tangential elastic force seems to be an origin of kinetic friction in this single-particle system.  However, if the rotational degree of freedom is taken into consideration, that is,  $\theta_i$ changes with time according to Eq.~(4), the average resistance force $F_0$ becomes zero for a sufficiently small $v$.  Figure 11(b) shows the numerical results at $h=1.68$ and 1.64 for $K_t=0.01K$ and $\eta_t=\eta$. 
Thus, we conclude again in our numerical simulation that no kinetic friction appears in a single-particle system.  

\section{Summary and Discussion}
We have studied frictional forces using a few inelastic particles intervening between two rough surfaces. 
The simplest model is a single-particle system without the rotational degree of freedom.  Even in this simplest system, the dynamical behavior is rather complex.  We have found a kind of jamming transition by changing the interval $h$ between the two walls. 
Below the critical interval, the intervening particle is trapped in the lower (or upper) wall.  Time-averaged resistance force $\langle F\rangle$ is roughly proportional to the constant velocity $v$ of the upper wall. 
No kinetic friction appears in a single-particle system, because $\langle F\rangle$ approaches zero in the limit $v\rightarrow 0$.    
At $h>h_c$, a chaotic time evolution is found, and the average resistance force  is approximated as  $\langle F\rangle\propto v^2$. 
We have confirmed that the maximum static friction is equal to the maximum value of the time-periodic resistance force $F(t)$.  The maximum static friction corresponds to the yield stress for the granular solidlike state. We have also confirmed the appearance of a stick-slip motion even in this single-particle system, when the upper wall is forced via a spring. 

Kinetic friction has appeared in a two-particle system. The mutual collisions between the intervening particles seem to play an important role in the appearance of kinetic friction and the nontrivial average velocity $V=(2/3)v$. 
At the moment of mutual collisions, impulsive forces are exerted on the lower and upper walls. The amplitude of the impulsive forces does not decrease even for  a sufficiently small $v$. The resistance force $F(t)$ behaves as $F(t)=F^{\prime}(vt)$ for a sufficiently small $v$; therefore, the time average of $F(t)$ takes a nonzero constant value even for a sufficiently small $v$.  We have further observed the propagation of mutual collisions in a ten-particle system.  The mutual collisions between the intervening particles occur even for a sufficiently small pulling velocity. The mutual collisions between intervening particles induce the dissipation of kinetic energy in the $x$ direction, which might lead to  kinetic friction; however, the detailed mechanism of this phenomenon is not yet completely understood theoretically.  

Our model is similar to M\"user's model in that the intervening particles play an important role. However, no kinetic friction appears if there is only one intervening particle, but kinetic friction appears as a result of the mutual inelastic collisions between the intervening particles. 
We consider that the origin of kinetic friction in our model is different from that in M\"user's model.

Finally, we have incorporated tangential viscous and elastic forces in the model equations. If the tangential viscous forces are included, the intervening particle exhibits a unidirectional motion with $V=(1/2)v$, when the tangential viscosity is larger than its critical value. However, the appearance or nonappearance of kinetic friction do not depend on the tangential viscous force. 
If both the tangential viscous force and tangential elastic force are included and the rotational degree of freedom is not taken into consideration, the nonzero resistance force $F_0$ appears. However, if the rotational degree of freedom is taken into consideration, $\langle F\rangle$ becomes zero for a sufficiently small $v$. 

Our model system is simple, but its dynamical behavior is rather complex. It is left for future study to analyze the system  from the viewpoint of the theory of dynamical system, as well as to associate the present results for a small number of particles with the dynamical behavior of many-particle systems.

\end{document}